\documentclass[prb,showpacs,superscriptaddress,twocolumn]{revtex4}
\usepackage{epsfig}
\usepackage{rotating}
\usepackage{graphicx}
\usepackage{amsmath}

\begin{document}

\title{c-axis transport in highly anisotropic metals: role of small
polarons}

\date{\today}

\author{A. F. Ho}
\affiliation{School of Physics and Astronomy, Birmingham University, 
Edgbaston, Birmingham B15 2TT, U.K.}

\author{A. J. Schofield}
\affiliation{School of Physics and Astronomy, Birmingham University, 
Edgbaston, Birmingham B15 2TT, U.K.}


\begin{abstract}

We show in a simple model of interlayer hopping of single electrons, that
transport along the weakly coupled  $c$-axis of quasi-two-dimensional metals
does not always probe only the in-plane electron properties. In our model
where there is a 
strong coupling between electrons and a bosonic mode that propagates in the
$c$ direction only, we find a broad maximum in the $c$-axis resistivity
at a temperature near the characteristic energy of the bosonic mode, while no
corresponding feature appears in the $ab$ plane transport. At temperatures
far from this bosonic energy scale, the  $c$-axis resistivity does track
the in-plane electron scattering rate. We demonstrate a reasonable
fit of our theory to the
apparent metallic to non-metallic crossover in the  $c$-axis resistivity
of the layered ruthenate  $\rm Sr_2 Ru O_4$.


\end{abstract}


\maketitle

Many metallic systems of current interest have highly anisotropic
electronic properties with one or more directions much more weakly
conducting than the others. Examples include the cuprate
metals~\cite{ito_1991a}, some ruthenates~\cite{tyler_1998a}, and many
organics~\cite{singleton_2000a}, where a measure of this anisotropy
is the resistivity ratio, $\rho_{c}/\rho_{ab}$,
which can be of order $10^3$ to $10^5$. (Henceforth, $c$ will denote
the weakly coupled direction with $a$ and $b$ the more conducting.)
Many of these materials have unusual metallic or superconducting
properties which have been attributed to strong correlation effects in
the $ab$ plane. Although the theoretical understanding of strongly
correlated electrons is incomplete, the electron anisotropy can be
used advantageously to give insights into the in-plane physics. By
assuming that the coupling up the $c$-axis occurs only by weak
single-electron tunneling, many
workers~\cite{kumar_1992a,mckenzie_1998a,sandeman_2001a} have shown
that $c$-axis resistivity depends {\em only} on the {\it in-plane} electron
Green's function.

This insight makes all the more puzzling the observation that in many
of these materials the resistivity-anisotropy ratio is strongly temperature
dependent with a cross-over from metallic behavior to insulating
behavior in the out-of-plane direction only. For example, ${\rm
Sr_2RuO_4}$~\cite{tyler_1998a} 
shows $T^2$ resistivity both in the plane and out-of-plane
for $T< 20$K but $\rho_c$ shows a broad maximum at around $130$K
and above this temperature the resistivity starts to decrease
 with increasing temperature (see Fig.~\ref{fitdata} below). 
In marked contrast, the in-plane resistivity shows
$\partial \rho/\partial T>0$ for all temperatures.
Evidently the low temperature metal is a 3D Fermi liquid, so this begs
the question how the out-of-plane transport is being blocked while
the in-plane transport remains metallic. In this paper we consider
how anisotropic coupling of the in-plane electrons to bosonic 
degrees of freedom
can give rise to
this effect. We show that if the electron-boson coupling is stronger
in the inter-plane direction then, at the characteristic energy scale
of the boson mode, there can be a maximum in the $c$-axis resistivity
with no corresponding transport feature for currents in the plane.

Other approaches to this puzzle have been offered previously. One
possibility is that the in-plane metal is a non-Fermi
liquid~\cite{anderson_1997a,vozmediano_2002a} 
or otherwise reflects unusual in-plane
physics~\cite{littlewood_1992a} such as superconducting
fluctuations~\cite{ioffe_1999b}. These approaches have generally been
argued to apply to the cuprate metals, but our motivation comes from
the layered strontium ruthenate compounds where the appeal to unusual
in-plane physics is less well grounded. Alternatively the $c$-axis
coupling may consist of more than single electron tunneling: with
disorder and boson-assisted hopping~\cite{rojo_1993a} or perhaps
inter-plane charge fluctuations~\cite{turlakov_2000a}. 

In this paper we also go beyond single particle hopping as the only
coupling between the planes and consider electrons coupled to a
bosonic mode.  However, in contrast to the above work, we use a
canonical transformation\cite{Firsov} to treat the electron-boson
coupling, and thus can deal with the strong coupling regime.  Our
assumptions are, first, that there is a well defined separation of
energy scales: $t_{ab} \gg \omega_0 \gg t_c$, where $t_{ab}$ is the
in-plane kinetic energy scale, $\omega_0$ is the characteristic energy
of the bosonic mode ({\it e.g.} the Debye temperature) and $t_c$ is
the single particle tunneling between layers. This assumption is
certainly well founded for ${\rm Sr_2 Ru O_4}$, as shown, for example,
in Ref.~[\onlinecite{Bergemann_2000}]. With this assumption it is
sufficient to compute conductivity to leading order in $t_c$.  Our
second assumption is that electron-electron correlations dominate over
the in-plane coupling to the bosonic mode such that the formation of a
small polaron {\it in the planes} is suppressed by the electronic
interactions. We will therefore be considering a limit where the
quasiparticle excitations of the system are insensitive to bosonic
modes whose wavelength in the plane is short compared to the size of a
quasiparticle (assumed to be large). With these assumptions we show
that anisotropic coupling can lead to differences between the in-plane
and out-of-plane conductivities that are consistent with the
experiments. A broad maximum can appear in the $c$-axis resistivity
$\rho_c$ as a function of temperature, while no such feature exists in
the in-plane resistivity $\rho_{ab}$.  Only away from the broad
maximum, does $\rho_c$ track the intrinsic in-plane scattering
rate. We shall also show that our model calculations can fit
reasonably the $\rm Sr_2 Ru O_4$ data of Tyler {\it et
al.}~\cite{tyler_1998a}, using physically plausible parameters.

Strictly our assumption that the electrons couple strongly only to
out-of-plane bosonic modes should be regarded as phenomenological, and
in this paper we compute the consequences of such an approach and
compare with experiment. Nevertheless we can offer some justification
for it.  In a quasi-2D metal, the anisotropy of the crystal structure
has important consequences for both the electronic structure and the
bosonic modes the electrons can couple to. Usually the anisotropy in
the characteristic energies of phonon modes are much smaller than
those than the electronic structure: the ruthenate that we will be
concerned with is thought to be an ionic (not covalent-bonded)
crystal. Where significant anisotropy can arise is in the strength of
the electron-phonon couplings for these different phonons.
Krakauer{\it et. al.} \cite{Krakauer_1993} have found in an LDA
calculation that the electron phonon coupling is much stronger when
atomic displacements are perpendicular to the Cu-O plane in LSCO,
while Kim {\it et. al.}\cite{Kim_1989} and Grilli and
Castellani\cite{Grilli_1994} found that strong in-plane electronic
correlation suppresses in-plane electron phonon coupling. Usually strong 
electron-phonon coupling leads to the well-studied small
polaron regime, where electronic motion becomes affected by
the accompanying large cloud of phonons\cite{Holstein,Firsov}. (For a
review, see [\onlinecite{mahan_1990a}].)
In this paper, our assumptions will lead to a small polaron forming
only between the layers. 

Our model Hamiltonian is then: 
\begin{eqnarray} 
\label{model}
H & = & \sum_{n} H^{(n)} + H_c + 
H_{\rm e-b} + H_{\rm b}  \; ,\nonumber \\
H_c & = & \int d^2 x \sum_n t_c c^{\dagger}_{n+1}(\vec{x}) c^{}_{n}(\vec{x}) 
+ {\rm H.c.} \; . 
\end{eqnarray}
The model system consists of a stack of 2D planes described by
$H^{(n)}$ containing arbitrary intra-planar interactions. In this work
it will be sufficient to characterize these planes by a
phenomenological 2D Green's function.  In $H_c$, electrons are assumed
to hop from one plane directly to a neighboring plane only, at the
same in-plane coordinate $\vec{x}$.  For simplicity, we have omitted
the spin index on the electron operators $ c_{n}(\vec{x})$. 
Many approaches~\cite{kumar_1992a,mckenzie_1998a,sandeman_2001a} 
assume only this coupling between the planes in a quasi-2D system. We go
beyond that by including a bosonic mode that also couples the
layers 
\begin{eqnarray}
\label{H-e-ph}
H_{\rm e-b}  & = & \sum_n \sum_q  M_q \exp(i q n c) \rho_n
 \left(a_q + a^{\dagger}_{-q}\right) \; .
\end{eqnarray} 
Our analysis can be applied to any gapped neutral bosonic mode with
long wavelength in the plane, such as a magnetic collective mode, but
we have in mind the (strong) coupling to a phonon mode with
displacement in the $c$ direction and long (infinite) wavelength in
the plane.  Here $q$ is along the $c$-axis, with $c$ the inter-layer
distance.  In this limit, only the total charge $\rho_n = \int d^2 x
c^{\dagger}_{n}(\vec{x}) c_{n}(\vec{x})$ of the $n$'th plane couples
to the boson creation and annihilation operators $a^{\dagger}_q, a_q$
so, {\it by construction}, these bosons do not dominate charge
transport within the plane.

Even with this limiting electron-phonon coupling, interlayer hopping
can affect in-plane correlations by spreading the in-plane correlation
from one plane to another, especially since the in-plane correlations
are supposed to be strong for a pure single layer in the cuprates and
also the ruthenates. However, this requires the interlayer hopping to
be coherent, which means the temperature must be low compared to
$t_c$: otherwise, the $c$-axis transport is diffusive and cannot
propagate correlations between the layers.  In this paper, we have
assumed a phenomenological form for the in-plane spectral weight, and
so we cannot address the issue of the strongly correlated {\it three}
dimensional state at $T \ll t_c$.

The Hamiltonian for the $c$-axis bosons, $H_{\rm b} = \sum_q \omega_q
a^{\dagger}_q a_q$, defines a dispersion $\omega_q$.  In this paper we
shall look closely at the Einstein phonon $\omega_q = \omega_0$. This
is both because optical phonons tend to have little dispersion
relative to acoustic ones, and also we are able to obtain analytic
results for this case. We shall also discuss, somewhat more
qualitatively, the opposite case of a generic dispersion where the
phonon density of states does not have any sharp features. For the
data fitting to $\rm Sr_2 Ru O_4$, we will use a simple form 
that allows us to test if the putative mode does indeed disperse. 

Of course electron-phonon systems have been well studied, but our
findings will be  shown to be clearly different to ones found 
in the small polaron models  studied in the
classic papers\cite{Holstein,Firsov,mahan_1990a}: small anisotropy
in electron-boson coupling cannot lead to the properties we shall
demonstrate below. In particular, at temperatures greater than the
temperature of the maximum in $\rho_c(T)$, our model has $\rho_c(T)$
tracking  the intrinsic in-plane scattering rate while in standard
small polaron theory (even if generalized to some degree of anisotropic
coupling), $\rho_c(T)$ keeps on decreasing (exponentially) with increasing
$T$. In short, in our model, the small polaron broad maximum in $\rho_c(T)$
is ``grafted'' on top of the background {\it in-plane} scattering.
We note that Kornilovitch\cite{kornilovitch_1999a}
has studied a similar anisotropic model: he did not calculate the
d.c. conductivity, and, with the effective mass approach employed, he
cannot access the non-metallic regime [beyond the broad maximum of
$\rho_c(T)$]. After completion of our theory, we became aware of the
 related work of Lundin and McKenzie~\cite{lundin_2002a} who
also studied a small polaron model for anisotropic metals. 
Their model is different to ours in an essential way: they take the
boson mode to be uncorrelated from one layer to another, while we have
a boson mode that propagate coherently along the $c$-axis.   

The key physics we are considering is the effect of strong
electron-phonon coupling on the charge transport in the weakest
direction, the $c$-axis. Physically, the motion of the electron is
accompanied by the emission and absorption of a large number of
phonons due to the strong coupling, forming the so-called small
polaron\cite{Holstein}. We treat 
$H_{\rm e-b}$ exactly by the canonical transformation, $\bar{H} =
\exp(-S) H \exp(S)$, a straightforward generalization of the
transformation well known in the small polaron
problem\cite{Firsov,mahan_1990a},
\begin{equation}
S = - \int d^2 x 
 \sum_{n, q} \frac{M_q}{\omega_q} e^{i q n c} 
c^{\dagger}_{n}(\vec{x}) c_{n}(\vec{x}) \left(a_q - a^{\dagger}_{-q}\right) .
\end{equation}
Then, $\bar{H} = \sum_{n} H^{(n)} + \bar{H}_c + H_{\rm b}$, and
\begin{eqnarray} 
\label{H_can}
\bar{H}_c & = &
- \sum_{n, q} \frac{|M_q|^2}{\omega_q} \rho_n \nonumber \\
 & + &\sum_n \int d^2x \; t_c c^{\dagger}_{n+1}(\vec{x}) c_{n}(\vec{x}) 
X^{\dagger}_{n+1} X_{n} 
+ {\rm H.c.} \; ,
\end{eqnarray}
\begin{equation}
{\rm where} \quad 
X_n = \exp \left\{ \sum_{q} \frac{M_q}{\omega_q} e^{i q n c}
 \left(a_q - a^{\dagger}_{-q}\right) \right\} \; . 
\end{equation} 
Thus strong electron-phonon coupling leads to (1) a renormalization
of the in-plane chemical potential (first term of $\bar{H}_c$), which
we shall henceforth ignore, and (2) an effective vertex correction for
the $c$-axis hopping $t_c$ (second term of $\bar{H}_c$).  At
temperatures lower than the characteristic energy of the boson,
electrons can hop coherently from one plane to another, with some
suppression due to the ``dragging'' of the boson cloud. With
increasing $T$, the hopping electron gets inelastically scattered by
more and more bosons: inter-layer hopping is now more diffusive.
Hence a crossover to non-metallic behavior occurs.
 
We consider the d.c. conductivity and the orbital
magnetoconductivity for fields in the $ab$ plane. 
For the conductivity, since the charge in the $n$'th
plane is $Q^{c}_n = e \int d^2 x c^{\dagger}_{n}(\vec{x}) c_{n}(\vec{x})$,
the current in the $c$-direction is just $j^{c}_n = \partial_t Q^{c}_n
= -i [ Q^{c}_n , H ]$. After the canonical transformation, 
$j^{c}_n = -i [ Q^{c}_n , \bar{H}_c ]$, and so:
\begin{eqnarray}
\left\langle j^{c}_n \right\rangle = i e t_c \int d^2 x 
\left\langle \left[
c^{\dagger}_{n+1}(\vec{x}) c_{n}(\vec{x}) X^{\dagger}_{n+1} X_{n} 
- {\rm h.c.}\right] \right\rangle \;.
\end{eqnarray}

The orbital effect of a magnetic field can be included using a Peierls
substitution~\cite{altland_1999a}.  The mathematical details of our
derivation follow our analysis reported in Ref.~\onlinecite{ho_2002a}.
Using the Kubo formula and expanding to $O(t_c^2)$ gives the
conductivity, {\it i.e.}, the linear response to an applied electric
field in the $c$-direction $\sigma_c= j^{c}/E^c$:
\begin{eqnarray} 
\sigma_c (B) & = & \lim_{\Omega \rightarrow 0} 
\frac{e ^2 t_c^2}{\Omega} 
\sum_n \int d^2 k_{\|} \int d \tau \nonumber \\
& & \times e^{i \Omega \tau} U(\tau)
G^{(2D)}_n(\vec{k}_{\|}, \tau) G^{(2D)}_{n+1}(\vec{k}_{\|}+\vec{q}_B, \tau)
\; ,
\label{sigma1}
\end{eqnarray}
\begin{equation}
{\rm where} \; U(\tau) = \left\langle X^{\dagger}_{n+1}(\tau) X_{n}(\tau) 
X^{\dagger}_{n}(0) X_{n+1}(0) \right\rangle_{H_{\rm b}} ,
\end{equation} 
$G^{(2D)}_n(\vec{k}_{\|})$ is the in-plane (dressed) Green's function
for plane $n$ at in-plane momentum $\vec{k}_{\|}$, and
$\vec{q}_B=e\vec{c} \times \vec{B}$ with $\vec{c}$ being the
inter-plane lattice vector. The temperature dependent factor $U(\tau)$ in
Eq.~\ref{sigma1} leads to the $c$-axis conductivity being no longer simply a
convolution of the in-plane electron Green's functions and we explore
the consequences in the following.

To make further progress we use a phenomenological 
in-plane electron Green's function in the spectral representation:
$G^{(2D)}_n(\vec{k}_{\|}, \omega_m) = \int dz
\frac{A^{(2D)}(\vec{k}_{\|}, z)}{i \omega_m - z}$, and $U(\nu_n) =
\int dz \frac{b(z)}{i \nu_n - z} $, then the Matsubara sums can be
done. We shall take the in-plane electron spectral function to have
the form: $ A^{(2D)}(\vec{k}, \omega) = \frac{1}{\pi}
\frac{\Gamma(\omega,T)}{(\omega - \epsilon_{\vec{k}})^2 +
\Gamma(T,\omega)^2}$, where we assume that the scattering rate 
$\Gamma(T,\omega)$ has no in-plane momentum dependence. Then,
integrating over $\vec{k}_{\|}$, the zero-frequency, zero-momentum
$c$-axis conductivity becomes:
\begin{eqnarray} 
\sigma_c(T,B) = \frac{e^2 N t_c^2}{\pi c T} 
 {\rm Re} \int d\omega d\nu \frac{ n_F(\omega)-n_F(\omega+\nu)}{\nu} && 
\nonumber \\
\times 
\frac{D(\nu,T)}{\left[(v_{\rm F} q_B)^2 -(\nu + i \Gamma
(T,\omega) + i \Gamma(T, \omega+\nu))^2\right]^{1/2}} &,&
\label{SIGMAr}
\end{eqnarray} 
where $D(\omega,T) = \omega 
b(\omega) n_B(\omega) \left[1+n_B(\omega)\right]$, $n_{F,B}$ are the
Fermi and Bose distributions and $N$ is the electron density. 
$D(\omega,T)$ can be 
calculated\cite{ho_2002a} following Ref.\onlinecite{mahan_1990a}. 
We now analyze this equation in detail for Einstein modes and, more
qualitatively, for a general dispersing mode.

For Einstein modes, $\omega_q = \omega_0$, the function $D(\omega,T)$
is made up of delta functions at the harmonics $\omega = n \omega_0,
\; n = 0, \pm1, \pm2, \ldots$\cite{mahan_1990a}.  It can be shown that
for experimental temperatures, one  needs only the first few harmonics.
We now discuss the three temperature regimes ($T\ll \omega_0, T\lesssim \omega_0,
T \gg \omega_0$) separately, and show that the low and high temperature
regimes basically tracks the in-plane scattering rate, just like in previous
studies of interlayer transport\cite{kumar_1992a,mckenzie_1998a,sandeman_2001a}.
Only in the regime where the temperature is near the boson scale $\omega_0$
can the electron-boson scattering dominate over the in-plane scattering
contribution, and instead, leads to a small polaron like broad maximum in
resistivity.   

At low temperatures ($T\ll \omega_0$), the asymptotic form for the
$c$-axis resistivity is the same as for band electrons, but with an
effective hopping parameter $t_c^{\rm eff} = t_c
e^{-\Delta^2/2\omega_0^2}$, where $\Delta^2 = \sum_q 2 |M_q|^2 (1-\cos
q)$ characterizes the strength of the electron-phonon interaction:
\begin{equation} 
\label{Einslo}
\sigma_c \simeq  \frac{ e^2 N t_c^2}{2 c \pi \Gamma(T,0)}
e^{-(\Delta/\omega_0)^2} \;\quad  (T \ll \omega_0) .
\end{equation}
For $T\gg \omega_0$, then $\sigma_c \sim T^{-\eta-\frac{1}{2}}$, using the
scattering rate $\Gamma(T) \propto T^{\eta}$. ($\eta=2$ for a Fermi liquid,
while for the marginal Fermi liquid, $\eta=1$.) Thus the low temperature
$B=0$ conductivity directly probes the in-plane scattering, while
the high temperature  conductivity still reflects the in-plane
scattering with an extra $T^{-1/2}$ 
factor~\cite{footnote2}.
  
\begin{figure}[h]
\includegraphics[width=\columnwidth]{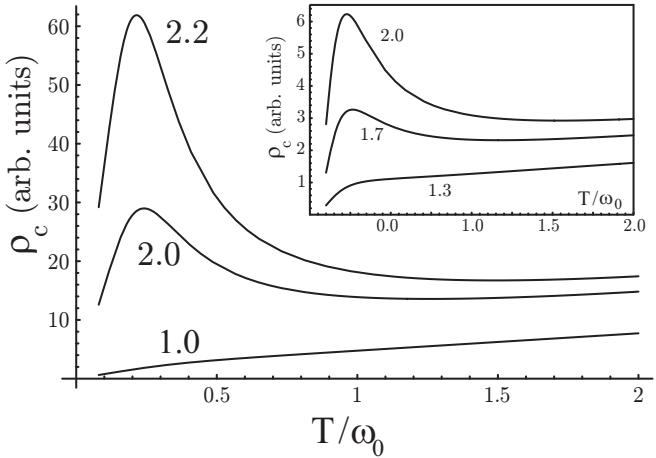}
\caption{The zero-field $c$-axis resistivity with in-plane marginal Fermi 
liquid scattering $\Gamma = \omega \coth\left(\omega/T\right)$ 
for electrons coupled to $c$-axis Einstein phonons for different
values of electron-phonon coupling $(\Delta/\omega_0)$. Inset: as
above but for a dispersing mode $\omega_q = \omega_0 [1- 0.4 \cos(q)] $.}
\label{inter}
\end{figure}

However, beyond a critical phonon coupling ($\Delta_c$) 
there is an intermediate temperature region where a broad maximum
appears in the $c$-axis resistivity at temperature $T_{\rm max}$ 
(Fig.~\ref{inter}). Beyond $T_{\rm max}$, the 
resistivity dips  to a (broad) minimum at $T_{\rm min}$ before
finally joining onto the asymptotic $T \gg \omega_0$ regime mentioned
above . 
We estimate, by using only the zeroth harmonics $n=0$, that if
the in-plane scattering rate has the form $\Gamma(T) = \alpha
T^{\eta}$, then the maximum appears when
\begin{equation} 
\Delta > \Delta_c \approx \omega_0\sqrt{\eta/0.4} \; .
\label{Delta_c}
\end{equation}
The position of the resistivity peak for $\Delta \agt \Delta_c$
is found to be
$ T_{\rm crit} \approx \omega_0 / 3.0 $ .
For $\Delta$ only slightly larger than $\Delta_c$, 
$T_{\rm max}$ and  $T_{\rm min}$ is given by: 
\begin{eqnarray}
\frac{\omega_0}{2 T_{\stackrel{\scriptstyle \rm max}{\rm min}}} 
\approx \frac{\omega_0}{2 T_{\rm crit}} \pm \sqrt{\frac{\eta}{0.2}
\left[\left(\frac{\omega_0}{\Delta_c}\right)^2 
-\left(\frac{\omega_0}{\Delta}\right)^2 \right]} . 
\end{eqnarray}
Note that both $T_{\rm max}$ and $\Delta_c$ depend mainly on phonon
parameters; one can show that the in-plane scattering rate enters only
in the form of the exponent $\eta$. Hence the width of the broad
maximum is governed by the scale $\omega_0$.

For a more general dispersing bosonic mode, $D(\omega,T)$ can be
calculated approximately\cite{mahan_1990a} assuming strong electron-phonon
interactions. Using the usual approximation
$[n_F(\nu)-n_F(\nu+\omega)]/\omega \simeq -\partial n_F/\partial \nu \simeq
\delta(\nu)$ valid at low $T$ when $\Gamma(T,\omega)$ varies slowly,
we find
\begin{eqnarray} 
\sigma_c(T,B) = \frac{e^2 N t_c^2}{\pi c T} 
e^{-C(T)} \int \frac{d\nu}{\sqrt{\pi \gamma^2(T)}} 
\frac{\nu}{\sinh \frac{\nu}{2 T}} && \nonumber \\
\times{\rm Re}
 \frac{\exp \left(-\frac{\nu^2}{4\gamma^2(T)}\right)} 
 {\left[(v_{\rm F} q_B)^2 - (\nu + 2 i
     \Gamma(T,\nu))^2\right]^{1/2}} \; ,&&
\label{siggen}
\end{eqnarray}
where
$$
C(T) = \sum_q F_q 
\frac{2 \sinh^2 \left(\frac{\omega_q}{4 T}\right)}
{\sinh\frac{\omega_q}{2 T}}, \quad 
\gamma^2(T) = \sum_q \frac{F_q |\omega_q|^2}{2 \sinh(\omega_q/2T)}\; ,
$$
and $F_q = \left|\frac{M_q}{\omega_q}\right|^2 2 (1-\cos q)$.

For illustration, we consider $\omega_q = \omega_0 [1- 0.4 \cos(q)] $,
and set the electron-phonon coupling $M_q$ to be
$q$-independent. Eq.~\ref{siggen} is then evaluated numerically and
plotted in the inset to Fig.~\ref{inter}.  $\bar{\Delta} = \sum_q F_q
\omega_q /2$ characterizes the strength of the electron-phonon
interaction in this case. The result is qualitatively similar to that of a
non-dispersing mode.

Now we consider the magnetic field dependence of the conductivity 
for the Einstein mode.
At low temperatures $T\ll \omega_0$ we find that the magnetoresistance
reflects the usual cross-over from quadratic to linear field
dependence~\cite{schofield_2000a} at a scale determined only by the
in-plane electron scattering rate
\begin{equation}
\Delta \rho_c = {[\rho_c(B)-\rho_c(0)]/\rho_c(0)} =  \sqrt{1 + [\omega_c /
2 \Gamma(T,0)]^2} \; ,
\end{equation}
where $\omega_c=v_F e c B$ is the cyclotron frequency.
At higher temperatures ($T \agt \omega_0$), 
the weak-field magnetoresistance is quadratic in field 
$ \Delta \rho_c \propto (\omega_c /\Gamma_{\rm eff})^2 $ ,
and defines a new scattering rate, $\Gamma_{\rm eff}$, depending on
both the electronic scattering and the phonon frequency. 
Taking into account only up to $n=1$ harmonics,
with $\Gamma_0=\Gamma(T,0)$, and $\Gamma_1=\Gamma(T,\omega_0)$, we find
\begin{equation}
\Gamma_{\rm eff}^2 \approx 
\frac{\frac{1}{2\Gamma_0} +\frac{4 \Gamma_1}{4 \Gamma_1^2 +\omega_0^2}}
{\frac{1}{16 \Gamma_0^3} +\frac{2 \Gamma_1 (4 \Gamma_1^2 - 3
\omega_0^2)}{(4 \Gamma_1^2 +\omega_0^2)^3}} \; .
\end{equation}
As for the zero field resistivity, the magnetoresistance for
the dispersing mode case is qualitatively similar to that
for the Einstein mode and is positive for all temperatures.

We now consider in more detail how this model might apply to real
materials and, in particular, the ruthenate systems. The 2D electronic
nature of these materials reflects the crystal structure and so it is
natural that the electron-phonon interaction should be
anisotropic. While there are to date no calculation for the electron-phonon
interaction parameters for the ruthenates, in the iso-structural
LSCO cuprate family, Krakauer {\it et. al.}\cite{Krakauer_1993} 
have calculated that there
is a strong electron-phonon coupling only for modes corresponding to
atomic displacements perpendicular to the Cu-O plane, (partly) because of 
weak screening of the resulting electric fields in this direction.
  Moreover it has been argued that in the perovskite
structure the coupling to c-axis vibrations is
enhanced~\cite{kornilovitch_1999a}. There exist optical phonons with
the appropriate symmetry for $c$-axis transport\cite{Braden} in $\rm
Sr_2 Ru O_4$, and experimentally the broad maximum in the $c$-axis
resistivity has been linked to a structural phase
transition\cite{jin_2001a} in $\rm Ca_{1.7} Sr_{0.3} Ru O_4$ at around
the broad maximum temperature.  Thus one should consider the
possibility of electron-phonon interaction affecting the $c$-axis
transport.  In particular, both the $\rm Sr_2 Ru O_4$ and $\rm
Ca_{1.7} Sr_{0.3} Ru O_4$ systems exhibit\cite{tyler_1998a,jin_2001a}
qualitatively this broad maximum structure found in our simple model,
in the $c$-axis resistivity near to their characteristic ($c$-axis)
phonon energy. 


\begin{figure}[h]
\includegraphics[width=\columnwidth]{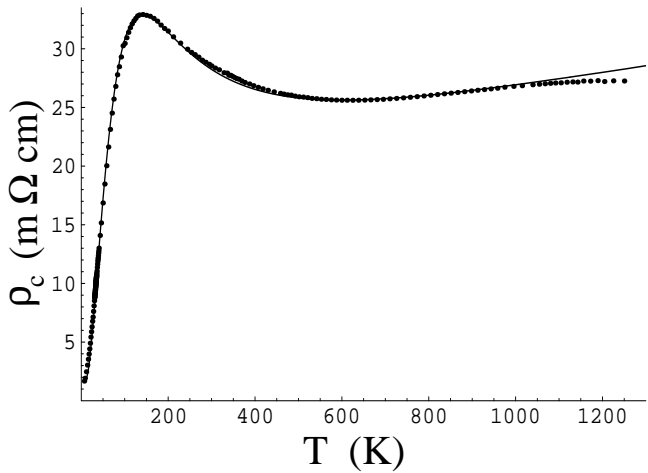}
\caption{Fitting the zero-field $c$-axis resistivity to the data of
Tyler {\rm et.al.}~\cite{tyler_1998a}. See the text for the explanation
of the theoretical curve (thin line) and the parameters used for the 
fitting. The data are plotted as points: for clarity, we have plotted only $3\%$
of the total number of available data points.}
\label{fitdata}
\end{figure}

A full quantitative modeling of $\rm Sr_2 Ru O_4$ is beyond the scope
of the simplified model presented here. 
Despite all these simplifying assumptions, our model does a reasonable fit
to the resistivity data of $\rm Sr_2 Ru O_4$ as we shall show, 
perhaps indicating some degree of
universality of the mechanism for c-axis transport studied in this paper.
Here for reference, we list the simplifications we have
assumed in our modeling:  1) the in-plane scattering rate can be
deduced directly from $\rho_{ab}$,  2) the main interlayer coupling
is single particle hopping from one layer to another and 3) a simple
phonon dispersion is employed (see later). 

For 1), this amounts to ignoring vertex corrections to in-plane
transport. In Fermi liquids, the vertex correction leads to an extra
cosine of the angle between in- and out-going electrons thereby correctly penalizing
back-scattering in conductivity. However, the qualitative trend is still
correct. As mentioned already, we have taken an in-plane spectral function
that has no in-plane momentum dependence. This ignores the complicated
multiple Fermi surfaces in  $\rm Sr_2 Ru O_4$. 
For 2), this means we ignore interlayer coulomb interaction. Also, we ignored the
multiband nature of $\rm Sr_2 Ru O_4$ and the dependence of $t_{\perp}$ on
in-plane momenta. For 3), we have taken the limiting case of an optical phonon mode
where the atomic displacements are perpendicular to the plane, and all the
atoms in the plane move together. We envisage that just as in LSCO, there will
be strong electron-phonon coupling only for phonons propagating mainly in the
$c$ direction\cite{Krakauer_1993}. Now in reality, because of the non-trivial
perovskite structure of $\rm Sr_2 Ru O_4$, these modes that can affect $c$-axis
transport will have relative atomic displacements both in-plane and out of plane.
But this makes electron hopping from one to another plane even more difficult,
as the polaron has to create disturbances both in-plane and out-of-plane. 

To some extent, these simplifying assumptions may only lead to some quantitative
changes in the fit parameters, because the calculation of $\rho_c$ (Eq.~\ref{sigma1})
involves an integral over in-plane momenta and thus averages out such dependences.

We now show in Fig.~\ref{fitdata} a fit to the $\rm Sr_2 Ru O_4$ 
data of  Tyler {\it et al.}~\cite{tyler_1998a}. The theoretical curve
is the thin continuous line, the data are the points. 
 The theoretical curve 
is generated as follows:  the in-plane scattering rate is approximated
as being proportional to the in-plane $\rho_{ab}$ (also taken from
the data of [~\onlinecite{tyler_1998a}]):
 $\Gamma(T,\omega) \approx A \left(\rho_{ab}(T) + \rho_0\right)$, 
where $A$ and $\rho_0$ are fitting parameters. We take a simple optical
boson dispersion $\omega(q) = \omega_0 \left(1- \lambda \cos q\right)$
with the boson bandwidth $\lambda$ and the characteristic energy
$\omega_0$ as fitting parameters. $\Gamma(T,\omega)$ and $\omega(q)$
are then fed into Eq.~\ref{siggen}, where the overall scale of
$\rho_c$ is found by fitting $\rho_c^{\rm max} \sim 33 $m$\Omega$ cm
to the peak of the theory curve. Despite the simplicity of the model
and the lack of knowledge of the exact boson dispersion form, the
theory curve is almost indistinguishable from the data points,
except at high $T$.\footnote{Other processes may intervene at
high temperature, but one source of discrepancy at high $T$ is that
the experimental data is the resistivity at constant {\it pressure}
while our calculation is for constant {\it volume}. Non-trivial
thermal expansion at higher $T$ will contaminate the data.}
The fitting is relatively insensitive to the precise values of the
parameters: some fine tuning (mainly, $\lambda$ and $\bar{\Delta}$)
is needed to get the positions of the
maximum, minimum and final upturn correctly.
 Furthermore, the parameters used are physically
plausible: $\lambda \approx 0.67, \omega_0 \approx 515$K,
$\bar{\Delta}/\omega_0 \approx  1.4, 
A/\omega_0 \approx 0.1 ($m$\Omega$ cm$)^{-1}, 
\rho_0 \approx 0.17$ m$\Omega$ cm. The value of $\omega_0$ is consistent
with the available phonon data of Braden {\it et.al.}\cite{Braden}, while the
size of the boson bandwidth suggests that the bosons do disperse in the
interplane direction. Note that the dispersion $\omega(q)$ always enter
the conductivity expression under a $q$-integral (see Eq.\ref{siggen}), 
and is the reason why
our simplified boson dispersion can still model the broad maximum
in the $c$-axis resistivity successfully. The size of the
electron-boson coupling parameter $\bar{\Delta}$ indicate that 
${\rm Sr_2RuO_4}$ is just slightly above threshold for obtaining
the maximum in $\rho_c$.
   

The $c$-axis magnetoresistance seen in ${\rm Sr_2RuO_4}$ is
unusual~\cite{hussey_1998a}, becoming negative for both transverse
($B\vert\vert ab$) and longitudinal ($B\vert\vert c$) fields above 80K
and maximally so at around 120K---coinciding with the peak
position. In this paper we have only considered orbital
magnetoresistance and, as might be expected, found it to be
positive~\cite{footnote}.  This is not inconsistent with the data
since the transverse magnetoresistance is always less negative than
the longitudinal one. If the change of sign is linked to the origin of
the resistivity maximum then, within the framework presented here, it
indicates that the frequency of, or coupling to, the bosonic mode is
field dependent.  At present, we have no microscopic picture of how
this might occur. 

In conclusion, we have shown that $c$-axis transport in a quasi-2D
metal does not always  probe only the in-plane electron properties.  Strong
coupling between electrons and a bosonic mode polarized in the 
$c$-direction in a highly anisotropic metal can lead to a broad
maximum---an apparent metallic to non-metallic crossover---in the
$c$-axis resistivity with no corresponding feature in the $ab$ plane
properties. The position in the temperature axis, and the shape of this broad
maximum are determined mainly by boson parameters and not on the magnitude of the
in-plane scattering rate. 
We have
discussed the potential application of the model to $\rm Sr_2 Ru O_4$
and its relative $\rm Ca_{1.7} Sr_{0.3} Ru O_4$. Despite certain
simplifying features of the model, the qualitative (and even quantitative)
properties of this
crossover in these layered ruthenates are captured succinctly.


   We are pleased to
acknowledge useful and stimulating discussions with M. Braden,
C. Hooley, V. Kratvsov, P. Johnson, M. W. Long, Y. Maeno, A. P. Mackenzie, 
G. Santi, I. Terasaki
 and Yu Lu. A.F.H. was supported by ICTP, Trieste where this work was
initiated, and by EPSRC (UK). A.J.S. thanks the Royal Society and the Leverhulme Trust
for their support.



\end{document}